\documentclass[12pt]{ronbun}

\usepackage{amsmath,amssymb}
\usepackage{cite}
\usepackage{bm}
\usepackage{dcolumn}

\newcommand{\nn}{\nonumber}

\newcommand{\del}{\delta}

\newcommand{\al}{\alpha}

\renewcommand{\th}{\theta}

\newcommand{\s}{\scriptscriptstyle}

\newcommand{\1}{\mathbb I}

\numberwithin{equation}{section}

\begin{document}

\begin{flushright}
\parbox{4.2cm}
{KEK-TH-904 \hfill \\
{\tt hep-th/0310035}
 }
\end{flushright}

\vspace*{1.1cm}

\begin{center}
 \Large\bf Dirichlet Branes of the   
Covariant Open Supermembrane \\
in AdS$_4~\times$ S$^7$ and AdS$_7~\times$ S$^4$
\end{center}
\vspace*{1.5cm}
\centerline{\large Makoto Sakaguchi$^{a}$ and Kentaroh Yoshida$^{b}$}

\begin{center}
\emph{Theory Division, High Energy Accelerator Research 
Organization (KEK),\\
Tsukuba, Ibaraki 305-0801, Japan.} 
\\
\vspace*{1cm}
$^{a}$Makoto.Sakaguchi@kek.jp
~~~~
 $^{b}$kyoshida@post.kek.jp
\end{center}

\vspace*{2cm}

\centerline{\bf Abstract}
  
\vspace*{0.5cm}
We discuss an open supermembrane theory in the AdS$_4\times S^7$ 
and AdS$_7\times S^4$ backgrounds. The possible Dirichlet branes of 
an open supermembrane are classified by analyzing 
the covariant Wess-Zumino term. 
All of the allowed configurations are related to those 
on the pp-wave background via the Penrose limit. 

\vfill

\thispagestyle{empty}
\setcounter{page}{0}

\newpage 

\section{Introduction}

Supermembrane theory \cite{BST,dWHN} is an interesting subject to study
in connection with the M-theory formulation \cite{BFSS}.  Open
supermembrane theory is also closely related to the Horava-Witten theory
\cite{HW}.  The Dirichlet $p$-branes of an open supermembrane in flat
space was discussed in Refs.\,\cite{EMM,dWPP}, where it was shown that only the
values $p=1,5$ and 9 are allowed. The $p=5$ case corresponds to the
M5-brane. The 9-brane implies the existence of the end-of-the world
9-brane as a boundary of the open supermembrane. On the other hand, it
is an interesting problem to consider boundary surfaces of open
supermembranes on the other backgrounds. We can take 
classical solutions of 
supergravities as consistent backgrounds for supermembranes \cite{BST}. 
That is, the kappa-invariance is ensured. 
In the recent progress, D-branes of the type IIB string on
the pp-wave background were studied in the covariant formulation
\cite{BPZ} by using the method of Lambert and West \cite{LW} (The same
method was applied to the type IIA string theory on a pp-wave
\cite{SY4} in Ref.\,\cite{HPS}). Moreover, we investigated the boundary
surfaces of an open supermembrane on the pp-wave background\footnote{
The matrix model on the maximally supersymmetric pp-wave background
\cite{KG} was proposed by Berenstein, Maldacena and Nastase 
in Ref.\,\cite{BMN}. This matrix model is closely related to the supermembrane
theory on the pp-wave background \cite{DSR,SY1,SY2}.  } in our previous
work \cite{SaYo:pp} by combining the method in Ref.\,\cite{LW} with that in
\cite{EMM,dWPP}, and we looked deeper into Dirichlet branes of an open
supermembrane on the pp-wave background more than in Ref.\,\cite{SY1}. In
particular, we found the 1/2 BPS configurations of 1-branes, M5-branes and
9-branes sitting at and outside the origin. 
The 9-brane configurations sitting at the
origin are intimately related to the heterotic matrix model on a pp-wave
\cite{M}.

In this paper we continue the study of Dirichlet branes of an open
supermembrane and consider the case of the open supermembrane in the
AdS$_{4/7}\times S^{7/4}$ backgrounds. By analyzing the covariant
supermembrane actions, the possible configurations of boundary surfaces
are classified. As the result, we find the configurations of 1-branes, 
M5-branes
and 9-branes preserving a half of supersymmetries at the origin.  The
AdS$_{4/7}\times S^{7/4}$ geometries are reduced to the Kowalski-Glikman
solution \cite{KG} via a certain Penrose limit \cite{P} as shown in
\cite{BFHP2}. Those are indeed related to the Dirichlet branes on the
pp-wave background via the Penrose limit and are consistent to our
previous results \cite{SaYo:pp}.

This paper is organized as follows: Section 2 is devoted to the setup
for our study. We introduce the supermembrane action in the
AdS$_{4/7}\times S^{7/4}$. In section 3, we investigate the Wess-Zumino
term in the covariant open supermembrane theory and classify the
possible configurations of Dirichlet branes of the membrane sitting at and
outside the origin. We also discuss the relation between the Dirichlet branes 
in the AdS$_{4/7}\times S^{7/4}$ and those on the pp-wave background. 
Section 4 is devoted to a conclusion and discussions.

\section{Covariant Action of Supermembrane on AdS$_4\times$ S$^7$
and AdS$_7\times$ S$^4$} 

In this section, we will introduce the action of the supermembrane
theory on the AdS$_4\times S^7$ and AdS$_7\times S^4$ background, and
briefly review its relevant aspects to our consideration.  We obtain the
covariant representation of the Wess-Zumino (WZ) term up to and
including fourth order in the $SO(10,1)$ spinor $\th$, which will be
analyzed in the next section.

The Lagrangian of the supermembrane \cite{BST,dWHN} is formally given as
a sum of the Nambu-Goto type Lagrangian $\mathcal{L}_0$ and WZ 
term $\mathcal{L}_{WZ}$\footnote{Our notation and convention are
summarized in Appendix.}
\begin{eqnarray}
&& \mathcal{L} \;=\;  \mathcal{L}_0
+ \mathcal{L}_{WZ}\,,\quad \mathcal{L}_0 \;=\;
- \sqrt{-g(X,\th)}  \,,
~~~
\mathcal{L}_{WZ} \;=\;\!B
\label{Lag}
\label{start}
\end{eqnarray}
where the induced metric $g_{ij}$ is the pull-back of the background
metric $G_{MN}$
\begin{equation}
 g_{ij} \; = \; E_i^M E^N_j G_{MN}
 =E_i^A E^B_j \eta_{AB}, \quad g \;=\; {\rm det}\,g_{ij}\,.
\end{equation}
and the three-form $B$ is defined by $H=dB$ with $H$ being the pull-back
of the four-form gauge superfield strength on the supergravity
background.  The term ${\cal L}_0$ is manifestly spacetime
superinvariant while ${\cal L}_{WZ}$ is quasi-superinvariant (i.e., 
superinvariant up to a surface term).  The $\kappa$-invariance of the
action must be imposed in order to match the fermionic and bosonic
degrees of freedom on the world-volume.  It is known that the condition
for the $\kappa$-invariance of the action is equivalent to the
supergravity equation of motion.  For an open supermembrane, the
$\kappa$-variation leads to a surface term, which must be deleted by
imposing appropriate boundary conditions on the boundary of the open
supermembrane.

We shall consider supermembranes on the AdS$_4\times S^7$
and AdS$_7\times S^4$
solutions of the eleven dimensional supergravity.
The supervielbeins of the AdS$_{4/7}\times
S^{7/4}$ backgrounds are \cite{dWPPS,C}
\begin{eqnarray}
& &
E^A \,=\, dX^M 
{e}^A_M
- i \bar{\th} \Gamma^A
\left(\frac{2}{\mathcal{M}}{\rm sinh}\frac{\mathcal{M}}{2}\right)^2
D\th\,, \quad
E^{\bar{\alpha}} \,=\, \left(
\frac{{\rm sinh}\mathcal{M}}{\mathcal{M}} D\th \right)^{\bar{\al}}\,, \\ 
& &
i\mathcal{M}^2 \; = \; 2(T_A{}^{B_1\cdots B_4}\th ) F_{B_1\cdots B_4}
(\bar{\th}\Gamma^A) \nn \\
&& \hspace*{2cm} - \frac{1}{288}
(\Gamma_{A_1A_2}\th)[
\bar{\th}(\Gamma^{A_1A_2B_1\cdots B_4}F_{B_1\cdots B_4}
+ 24\Gamma_{B_1B_2} F^{A_1A_2B_1B_2} 
)]\,, \nn\\ 
&& (D\th)^{\bar{\alpha}} 
\,\equiv\, d\th^{\bar{\alpha}} 
+{e}^A
(T_A{}^{B_1\cdots B_4}\th)^{\bar{\alpha}}
 F_{B_1\cdots B_4} 
- \frac{1}{4}{\omega}^{A_1A_2}
(\Gamma_{A_1A_2}\th)^{\bar{\alpha}}\,, 
\nn\\
& & T_A{}^{B_1\cdots B_4} \; = \; \frac{1}{288}
(\Gamma_A{}^{B_1\cdots B_4} - 8 \delta_A^{[B_1} \Gamma^{B_2B_3B_4]} )\,, \quad 
{e}^A
\,=\, dX^M {e}^A_M\,,\nonumber\\
&&F_{\mu_1\cdots\mu_4}=6f(\det{e^{A}_{M}})\epsilon_{\mu_1\cdots\mu_4}\,. 
\nonumber
\end{eqnarray}
When $f$ is pure imaginary (real),
we are dealing with AdS$_4\times S^7$ (AdS$_7\times S^4$). 
If we take $f\rightarrow 0$ limit, we can recover the flat Minkowski
spacetime. 
The vielbeins ${e}^A$ and spin connections ${\omega}^{AB}$
of the AdS$_{4/7}\times S^{7/4}$ backgrounds
can be computed by 
parameterizing the group manifold
corresponding to the algebra given in \cite{dWPPS} 
as $g={\rm e}^{X^aP_a+X^{a'}P_{a'}}$
and defining Cartan one-forms as 
\begin{eqnarray}
g^{-1}dg &=& e^a P_a + e^{a'}P_{a'}
+\frac{1}{2}\omega^{ab}J_{ab}+\frac{1}{2}\omega^{a'b'}J_{a'b'}\,, \\
e^A&:&
e^a=\left(\frac{\sinh Y}{Y}\right)^a{}_b \, dX^b,~~~
e^{a'}=\left(\frac{\sinh Y'}{Y'}\right)^{a'}{}_{b'} \, dX^{b'}\,,\\
\omega^{AB}&:&
\omega^{ab}=4f^2X^{[a}\left(
\frac{2}{Y}\sinh\frac{Y}{2}
\right)^2{}^{b]}{}_c \, dX^c\,, \\
&& \omega^{a'b'}=-f^2X^{[a'}\left(
\frac{2}{Y'}\sinh\frac{Y'}{2}
\right)^2{}^{b']}{}_{c'}\, dX^{c'}\,,
\end{eqnarray}
where
\begin{eqnarray}
(Y^2)^a{}_b=\sqrt{|4f^2|}((X^c)^2\delta^a_b-X^aX_b)\,, \quad 
({Y'}^2)^{a'}{}_{b'}=-\sqrt{|f^2|}((X^{c'})^2\delta^{a'}_{b'}-X^{a'}X_{b'})\,.
\end{eqnarray}
The metric of the background in this parameterization is obtained as  
$ds^2=\eta_{AB}e^Ae^B$. 

On the other hand, 
the third rank tensor $B$ of AdS$_{4/7}\times S^{7/4}$ is given by
\begin{eqnarray}
B &=& B^{(1)} + B^{(2)}\,, \\
B^{(1)} &=& \frac{1}{6}e^{A_1}\wedge e^{A_2} \wedge 
e^{A_3}\, C_{A_1A_2A_3}\,, \\
B^{(2)} &=&  i \! \int^1_0\!\!\! dt\, \bar{\th}\Gamma_{AB}E(x,t\th)\wedge 
E^A(x,t\th)\wedge E^B(x,t\th)\,,
\label{ads-WZ}
\end{eqnarray}
where $E^A(x,t\th)$ and $E^{\bar{\al}}(x,t\th)$ is obtained by 
shifting 
$\th \rightarrow t\th$ in $E^A$ and $E^{\bar{\alpha}}$.
The second WZ term $B^{(2)}$ becomes
\begin{eqnarray}
B^{(2)}&=&
\frac{i}{2}\bar\theta\Gamma_{AB}D\theta \wedge
dX^Me_M^A \wedge
dX^Ne_N^B
+\frac{1}{2}\bar\theta\Gamma_{AB}D\theta \wedge
\bar\theta\Gamma^AD\theta \wedge
dX^Me_M^B
\nonumber\\&&
+\frac{i}{24}\bar\theta\Gamma_{AB}{\cal M}^2D\theta \wedge
dX^Me_M^A \wedge
dX^Ne_N^B +{\cal O}(\theta^6)\,.
\end{eqnarray}
We do not need to expand the first term $B^{(1)}$ in the WZ term 
and the Nambu-Goto part ${\cal L}_0$ with respect to $\theta$,
because these terms turn out to be irrelevant in our analysis.
The covariant WZ term constructed above will be used in the
next section for the classification of boundary surfaces of an open
supermembrane on AdS$_{4/7}\times S^{7/4}$.

\section{Classification of Supersymmetric Dirichlet Branes}

Here we will classify possible supersymmetric Dirichlet branes of an
open supermembrane on the AdS$_{4/7}\times S^{7/4}$ background by using
the covariant action. In conclusion, we will find supersymmetric
1-branes, M5-branes and 9-brane sitting at the origin as Dirichlet branes of an
open supermembrane on AdS$_{4/7}\times S^{7/4}$.

\subsection{Boundary Conditions of Open Supermembrane}

To begin with, let us introduce the boundary condition on the boundary
$\partial\Sigma$ of the world-volume $\Sigma$ of an open supermembrane.
The coordinates should be distinguished in terms of boundary conditions,
and so we classify the coordinates of a Dirichlet $p$-brane as follows:
\begin{eqnarray}
&&\mbox{} \left\{
\begin{array}{l}
\overline{A}_i~~(i=0,\ldots,p)~~:~~ \mbox{Neumann directions} \quad 
\partial_{\bf n}X^{\overline{A}}\equiv
 \partial_{\bf n}X^{{M}}e_{M}^{\overline{A}} = 0 
\\
\underline{A}_j~~(j=p+1,\ldots,10)~~:~~ \mbox{Dirichlet directions}
\quad
\partial_{\bf t}X^{\underline{A}}\equiv
 \partial_{\bf t}X^{{M}}e_{M}^{\underline{A}} = 0 
\end{array}
\right.\,,
\end{eqnarray}
where we have introduced a normal vector $n^a$ to the boundary
$\partial\Sigma$, and the normal and tangential derivatives on the
boundary defined by
\begin{eqnarray}
\partial_{\bf n} \equiv n^a\partial_a\,, \quad \partial_{\bf t} \equiv
\epsilon^{ab}n_a\partial_b \quad (a,b=1,2)\,.
\end{eqnarray} 
Now we shall impose the boundary condition on the 
fermionic variable $\th$: 
\begin{eqnarray}
\label{p-} P^-\th |_{\partial \Sigma} = 0
\quad \mbox{or} \quad P^+\th |_{\partial \Sigma} = 0\,,
\end{eqnarray}
by using the projection operators $P^{\pm}$ defined as
\begin{eqnarray}
P^{\pm} \equiv \frac{1}{2}\left( \1\pm s{\rm M}^{10-p} \right)\,, 
\quad {\rm M}^{10-p}
\equiv  
\Gamma^{\underline{A}_{p+1}} \Gamma^{\underline{A}_{p+2}} \cdots
\Gamma^{\underline{A}_{10}}\,.
\end{eqnarray}
We choose $s=1$ when the time-direction $X^0$ is a Neumann direction, 
while $s=i$ when it is a Dirichlet direction.  
The requirement that $P^{\pm}$ are projection operators restricts the
value $p$ to $p=1,2,5,6$ and $9$.

From now on, we examine the $\kappa$-invariance of the action
(\ref{Lag}) on AdS$_{4/7}\times S^{7/4}$.  Under the $\kappa$-variation
\begin{eqnarray}
\delta_\kappa E^A=0~~~ \to ~~~ \delta_\kappa
X^M=i\bar\theta\Gamma^M\delta_\kappa\theta ~+{\cal O}(\theta^4)\,,
\label{kappa}
\end{eqnarray}
the action leads to a surface term only because non-surface terms vanish
for the supergravity solution.  The kinetic term ${\cal L}_0$ does not
contribute to the surface term.  The variation of ${\cal L}_0$ includes
$\delta_\kappa E_i^A=\partial_i(\delta_\kappa X^{\hat M})E_{\hat M}^A$
with $\hat M=(M,\alpha)$, but the surface term vanishes because
$\delta_\kappa X^{\hat M}E_{\hat M}^A=\delta_\kappa E^A=0$.  On the
other hand, we can take the vanishing surface term of the variation of
$B^{(1)}$ by having additional degrees of freedom at the boundary.  It
is thus enough to investigate the surface term of the $\kappa$-variation
of the Wess-Zumino term $B^{(2)}$.

It is convenient to divide the action $S_{WZ}^{(2)}=\int B^{(2)}$ into
the following three parts up to and including fourth order in $\theta$
\begin{eqnarray}
S^{f}&=&
\int\!\! d^3\sigma\, \epsilon^{ijk}\Bigl[
\frac{i}{2}\bar\theta\Gamma_{AB}
(
\partial_i\theta
+\partial_iX^CT_C{}^{a_1\cdots a_4}\theta~F_{ a_1\cdots a_4}
)
\partial_jX^A\partial_kX^B
\\&& ~~
+\frac{1}{2}\bar\theta\Gamma_{AB}
(
\partial_i\theta
+\partial_iX^CT_C{}^{ a_1\cdots a_4}\theta~F_{ a_1\cdots a_4}
)~
\bar\theta\Gamma^A
(
\partial_j\theta
+\partial_jX^DT_D{}^{ a_1\cdots a_4}\theta~F_{ a_1\cdots a_4}
)~
\partial_k X^B
\Big],\nonumber\\
S^{spin}&=&
-\frac{1}{4}
\int\!\! d^3\sigma\, \epsilon^{ijk}\Bigl[
\frac{i}{2}\bar\theta\Gamma_{AB}\Gamma_{D_1D_2}\theta
~\omega_C^{D_1D_2}
\partial_iX^C
\nonumber\\&&~~~~
+\frac{1}{2}\bar\theta\Gamma_{CA}\Gamma_{D_1D_2}\theta~\omega_B^{D_1D_2}~
\bar\theta\Gamma^C
 (\partial_i\theta+T_E^{~ a_1\cdots a_4}\theta F_{ a_1\cdots a_4}\partial_iX^E)~
\nonumber\\&&~~~~
-\frac{1}{2}\bar\theta\Gamma_{CA}
 (\partial_i\theta+T_D^{~ a_1\cdots a_4}\theta F_{ a_1\cdots a_4}\partial_iX^D)~
\bar\theta\Gamma^{C}\Gamma_{E_1E_2}\theta~\omega_B^{E_1E_2}
\nonumber\\&&~~~~
-\frac{1}{8}\bar\theta\Gamma_{CA}\Gamma_{D_1D_2}\theta~
\omega_B^{D_1D_2}~
\bar\theta\Gamma^C\Gamma_{E_1E_2}\theta~
\omega_F^{E_1E_2}~
\partial_iX^F
\Bigr]~
\partial_jX^A
\partial_kX^B,
\\
S^{\cal M} &=&
\frac{i}{24}
\int \!\! d^3\sigma\, \epsilon^{ijk}
\bar\theta\Gamma_{AB}{\cal M}^2D_i\theta~
\partial_jX^A
\partial_kX^B\,.
\end{eqnarray}
The third part $S^{{\cal M}}$ contains the $\mathcal{M}^2$ term, while
the second part $S^{spin}$ includes the spin connection. The first part
$S^{f}$ contains all terms other than $S^{{\cal M}}$ and $S^{spin}$.

We analyze these terms in turn. Because the $\kappa$-variation boundary
terms do not cancel out each other, we examine these terms separately.
First, we consider the $\kappa$-variation of $S^{f}$ and study the
configuration of Dirichlet branes sitting at the origin. As we will see
later, the other terms $S^{spin}$ and $S^{\cal M}$ do not affect the
result at the origin. We next investigate the configuration of Dirichlet
branes sitting outside the origin. In this case, the term $S^{spin}$
leads to additional conditions.

\subsection{Dirichlet Branes at the Origin}

Let us consider the $\kappa$-variation of $S^{f}$.
The surface terms come from the variation of variables with a derivative.
Under the variation (\ref{kappa}), the action leads to surface
terms up to and including fourth order in $\th$
\begin{eqnarray}
\label{S^f}
\delta_\kappa S^{f}&=&
-\int\!\! d\tau\!\!\int_{\partial \Sigma}\!\!\! d\xi\,\Bigl[\,
i\bar\theta\Gamma_{\overline{AB}}\delta_\kappa\theta
\dot X^{\overline{A}}\partial_{\bf t} X^{\overline{B}} \\
&& \qquad 
+\frac{1}{2}(
\bar\theta\Gamma^{{A}}\delta_\kappa\theta~
 \bar\theta\Gamma_{A\overline{B}}
+\bar\theta\Gamma_{A\overline{B}}\delta_\kappa\theta~
\bar\theta\Gamma^A
)
(
\dot{\theta}\partial_{\textbf{t}}X^{\overline{B}}
-
\partial_{\textbf{t}}\theta \dot{X}^{\overline{B}}
)
\nonumber\\ 
&& \qquad -\frac{1}{2}(
\bar\theta\Gamma_{\overline{AB}}~T_{\overline{C}}^{~ a_1\cdots a_4}
 \theta ~\bar\theta\Gamma^{\overline{C}}\delta_\kappa\theta
+\bar\theta\Gamma_{\overline{CA}}~T_{\overline{B}}^{~ a_1\cdots a_4}
 ~\bar\theta\Gamma^{\overline{C}}\delta_\kappa\theta
\nonumber\\
&& \qquad +\bar\theta\Gamma_{\overline{CA}}\delta_\kappa\theta~
 \bar\theta\Gamma^{\overline{C}}~T_{\overline{B}}^{~ a_1\cdots a_4}\theta
)~F_{ a_1\cdots a_4}~
(
\dot X^{\overline{A}}\partial_{\bf t} X^{\overline{B}}
-\dot X^{\overline{B}}\partial_{\bf t} X^{\overline{A}}
)
\,\Bigr]\,,
\end{eqnarray}
where a dot on a variable means the world-volume time derivative
$\partial_\tau$ of the variable.  The first line vanishes under the
condition:
\begin{eqnarray}
\bar\theta\Gamma_{\overline{AB}}\delta_\kappa\theta=0\,,
\end{eqnarray}
under which the second line also vanishes. This condition restricts the
value of $p$ to $1,~5$ and $9$. Thus we have rederived the well-known
result in flat space \cite{EMM,dWPP}.  It should be noted that only the first
and second lines survive in the flat limit $f\rightarrow 0$.

In order for the third and fourth lines to vanish, we must impose
the additional constraints: 
\begin{eqnarray}
\bar{\th}\Gamma_{\overline{A}\overline{B}}T_{\overline{C}}{}^{ a_1\cdots a_4}\th = 
\bar{\th}\Gamma^{\underline{C}}T_{\overline{B}}{}^{ a_1\cdots a_4}\th = 0\,.
\end{eqnarray}
These conditions are satisfied under one of the following
boundary conditions on
$( a_1, a_2, a_3, a_4)$-directions: 
\begin{itemize}
\item One of $( a_1,\cdots, a_4)$ is a Dirichlet direction and 
other three directions are Neumann directions. 
\item Three of $( a_1,\cdots, a_4)$ are Dirichlet directions and 
the remaining one direction is a Neumann direction. 
\end{itemize}
That is, the directions in which 9-branes, M5-branes and 1-branes can
span are restricted.  We shall classify the possible Dirichlet branes of
an open supermembrane on AdS$_{4/7}\times S^{7/4}$ below.  

First, let us consider the configurations of 9-branes, which is
classified as follows:

\vspace*{0.2cm}
$\bullet$~~\underline{{\bf Classification of 9-branes}} (at the origin)
\quad ($\sharp D =1$ and $\sharp N = 10$)
  \begin{enumerate}
   \item 
$(3,7)$-brane; ${\rm M}^1 = \Gamma^{ a}$\,.
  \end{enumerate}
The notation of $(3,7)$-brane means a brane with the world-volume
spanned along three directions of AdS$_4$ ($S^4$) and seven directions
of $S^7$ (AdS$_7$) for AdS$_{4}\times S^7$ (AdS$_{7}\times S^4$). 
The spanned directions satisfy the Neumann conditions, and the other
directions are the Dirichlet ones which are set to zero in the present
case. 

Next, we shall classify the configuration of M5-branes preserving a half
of supersymmetries at the origin. The following two types of M5-brane
configurations are allowed. 

\vspace*{0.2cm}
$\bullet$~~\underline{{\bf Classification of M5-branes}} (at the origin) \quad 
($\sharp D =5$ and $\sharp N = 6$)
   \begin{enumerate}
    \item $(3,3)$-brane; ${\rm M}^{5} = \Gamma^{ a a_1'\cdots  a_4'}$\,, 
    \item $(1,5)$-brane; ${\rm M}^{5} = \Gamma^{ a_1\cdots  a_3 a_1' a_2'}$\,.
   \end{enumerate}

In the end, we consider the 1/2 BPS configurations of 1-branes at the
origin. The possible configurations of 1-brane are given as follows:

\vspace*{0.2cm}
$\bullet$~~\underline{{\bf Classification of 1-branes}} (at the origin) \quad  
($\sharp D =9$ and $\sharp N = 2$)
   \begin{enumerate}
    \item $(1,1)$-brane; ${\rm M}^{9}=\Gamma^{ a_1 a_2 a_3 a_1'\cdots
	  a_6'}$\,.
   \end{enumerate}

Finally, we shall summarize the above result in Tab.\,\ref{cl:tab}. 
We would like to remark again that 
the Dirichlet branes sitting at the origin has been completely
classified, since, as explained below, 
the $S_{\mathcal{M}}$ and $S^{spin}$ have 
no effect on the above result at the origin. 
Moreover, it should be noted that our result agrees with the 
possible AdS embedding. 

\begin{table}
 \begin{center}
  \begin{tabular}{|c|c|c|c|}
    \hline
    & $p=9$  & $p=5$   & $p=1$    \\
    \hline  \hline
  ${\rm M}^{10-p}$     &$\Gamma^{ a}$ &
  $\Gamma^{ a a_1'\cdots  a_4'},~~\Gamma^{ a_1 a_2 a_3 a_1' a_2'}$ &
  $\Gamma^{ a_1 a_2 a_3 a_1'\cdots  a_6'}$   \\
    \hline
  \end{tabular}\caption{Classification of 1/2 BPS Dirichlet branes sitting at the
  origin.}
  \label{cl:tab}
 \end{center}
\end{table}

\subsection{Classification of Dirichlet Branes outside the Origin}

Here we will consider the contribution of the spin connection.
The variation of $S^{spin}$
is given by
\begin{eqnarray}
\del_{\kappa}S^{spin} &=& 
\frac{1}{4}\int\!\! d\tau\int_{\partial\Sigma}\!\!\!d\xi\,
\Bigl[
-\frac{1}{2}\bar\theta\Gamma_{\overline{AB}}\Gamma_{D_1D_2}\theta\cdot
\bar\theta\Gamma^{\overline{C}}\delta_\kappa\theta~\omega^{D_1D_2}_{\overline{C}}
-\frac{1}{2}\bar\theta\Gamma_{\overline{C}\overline{A}}\Gamma_{D_1D_2}\theta\cdot
\bar\theta\Gamma^{\overline{C}}\delta_\kappa\theta~\omega^{D_1D_2}_{\overline{B}}
\nonumber\\&&~~~~~~
-\frac{1}{2}\bar\theta\Gamma_{\underline{C}\overline{A}}\delta_\kappa\theta\cdot
\bar\theta\Gamma^{\underline{C}}\Gamma_{D_1D_2}\theta~\omega^{D_1D_2}_{\overline{B}}
\Bigr]~
(\dot X^{\overline{A}}\partial_{\bf t}X^{\overline{B}}
-\dot X^{\overline{B}}\partial_{\bf t}X^{\overline{A}}) + \mathcal{O}(\th^6)\,,
\end{eqnarray}
where the expression of the spin connection $\omega^{AB}_{{C}}=
e_{{C}}^M\omega_M^{AB}$ is given by 
\begin{eqnarray}
\omega^{AB}_C~~:~~
\omega^{ab}_c=-4f^2X^{[a}\left(
\frac{2}{Y}\tanh\frac{Y}{2}
\right)^{b]}_c,~~~
\omega^{a'b'}_{c'}=f^2X^{[a'}\left(
\frac{2}{Y'}\tanh\frac{Y'}{2}
\right)^{b']}_{c'}\,.
\end{eqnarray}
Let us consider the vanishing condition for the surface term.
The first two terms vanish when
\begin{eqnarray}
\bar\theta\Gamma_{\overline{AB}}\Gamma_{\overline{D}\underline{E}}\theta~
X^{\underline{E}}=0\,,
\end{eqnarray}
which is trivially satisfied for the branes sitting at the origin
$X^{\underline{A}}=0$.
On the other hand, this condition is satisfied only for $p=1$ in the
case of branes sitting outside the origin. 
The vanishing condition for the third term
\begin{eqnarray}
\bar\theta\Gamma^{\underline{C}}\Gamma_{\underline{A}\overline{D}}\theta~
X^{\underline{D}}=0\,,
\end{eqnarray}
is satisfied for branes sitting at the origin,
but allows only the 9-brane configuration if we consider outside the origin.
Thus, we have found that there is no solution of the 1/2 BPS Dirichlet branes
outside the origin. 

\subsubsection*{No Contribution from the $S^{\mathcal{M}}$}

Finally we shall discuss the contribution of $S^{\mathcal{M}}$ to 
our result. 
We can easily show that $S^{\mathcal{M}}$
including $\mathcal{M}^2$ does not change the above classification.
The variation of $S^{\mathcal{M}}$ is given by
\begin{eqnarray}
\delta_\kappa S^{\cal M} =
-\int\!\! d\tau\!\int_{\partial\Sigma}\!\!\! d\xi~
\frac{i}{12}\bar\theta\Gamma_{\overline{AB}}{\cal M}^2\delta_\kappa\theta~
\dot X^{\overline{A}}\partial_{\bf t}X^{\overline{B}},
\end{eqnarray}
and hence, in order for the surface terms to vanish, we need to impose
the additional condition: 
\begin{equation}
\bar{\th}\Gamma_{\overline{AB}}\mathcal{M}^2\del_{\kappa}\th = 0\,.
\end{equation}
By the use of the expression of $\mathcal{M}^2$, this condition can be
rewritten as
\begin{eqnarray}
&&
\Bigl(
2\bar{\th}\Gamma_{\overline{A}\overline{B}}T_{\overline{C}}{}^{ a_1\cdots a_4}\th\cdot
\bar{\th}\Gamma^{\overline{C}}\del_{\kappa}\th
- \frac{1}{288}\bar{\th}\Gamma_{\overline{A}\overline{B}}
\Gamma_{CD}\th\cdot\bar{\th}\Gamma^{CD a_1 a_2 a_3 a_4}
\del_{\kappa}\th \nn \\
&& \hspace*{6.5cm}
- \frac{24}{288}
\bar{\th}\Gamma_{\overline{A}\overline{B}}\Gamma^{a_1a_2}\th\cdot
\bar{\th}\Gamma^{a_3a_4}\del_{\kappa}\th
\Bigr)F_{ a_1\cdots a_4}= 0\,.
\label{cond-re}
\end{eqnarray}
The condition (\ref{cond-re}) is satisfied for the configurations 
in Tab.\,\ref{cl:tab}, as we can easily check. Because there is no
nontrivial 1/2 BPS configurations of Dirichlet branes sitting outside
the origin, we do not worry about the outside origin case. 
Therefore we now can say that the Dirichlet branes have been completely
classified.  

In this paper we have considered the $\kappa$-variation boundary terms
up to and including fourth order in $\theta$.  We expect that our
classification of Dirichlet branes is true even at the higher order in
$\theta$.  In fact, there are some arguments~\cite{BPZ} which support 
that this is the case.

\subsection{Penrose Limit of Dirichlet Branes in the AdS Background}

Now we will consider the Penrose limit of the Dirichlet branes of the
open supermembrane in the AdS$_{4/7} \times S^{7/4}$ backgrounds. 
By the use of the Penrose limit, the AdS$_{4/7} \times S^{7/4}$
geometries are reduced to the maximally
supersymmetric pp-wave background, which is often called 
Kowalski-Glikman (KG) solution. 
The KG solution can be obtained from AdS$_{4/7}\times S^{7/4}$ as the
Penrose limit only if one makes the light cone coordinates $X^\pm$ by taking
$X^0$ and $X^{10}$ from AdS$_{4/7}$ and $S^{7/4}$,
respectively. If we choose both coordinates $X^0$ and $X^{10}$
from AdS space or sphere, then the Penrose limit is trivial and we
obtain the eleven-dimensional flat Minkowski spacetime
\cite{BFHP2}. In this case, as a matter of course, all kinds of
9-branes, M5-branes and 1-branes are allowed.

Now we will briefly explain how to take the Penrose limit. To begin
with, we define the light-cone
projection operators as ${\cal P}_\pm= -
\frac{1}{2}\Gamma_\pm\Gamma_\mp$ and decompose the original spinor $\th$ 
as $\th = 1\cdot \th = - \frac{1}{2}\left(\Gamma_+\Gamma_- +
\Gamma_-\Gamma_+\right)\th = \th_+ + \th_-$. After rescaling
 the fermionic coordinates
$\theta_\pm \equiv {\cal P}_\pm\theta$ as $\theta_-\to\Omega\theta_-$
and $\theta_+\to\theta_+$, we take 
the limit $\Omega\to 0$. This is the definition of the Penrose limit in
our case.   

We will study the Penrose limit of each type of the possible 
Dirichlet branes at the origin in the AdS coordinate system. 
For simplicity, we examine the boundary condition 
$P^-\theta |_{\partial\Sigma}=0$ below, but the Penrose limit
of the condition $P^+\theta |_{\partial\Sigma}=0$ leads to the similar
result.

\subsubsection*{Penrose Limit of 9-branes} 

Now let us examine the Penrose limit of 9-brane configurations which 
is the $(3,7)$-brane.  The boundary condition is given by 
\begin{eqnarray}
s\Gamma^ a\theta=\theta\,.
\label{boundary:(3,7)}
\end{eqnarray}
When one of $1, 2, 3$ is the Dirichlet direction, the
condition (\ref{boundary:(3,7)}) reduces to $\Gamma^I\theta=\theta$,
$I=1,2,3$, under the Penrose limit.  This is nothing but the boundary
condition for $(+,-;2,6)$-brane with ${\rm M}^1=\Gamma^I$,
which is the possible configuration of 9-branes
in the pp-wave case~\cite{SaYo:pp}. 
Note that the resulting pp-wave background after the
Penrose limit is not described in the Brinkmann coordinates, but the
result of Ref.\,\cite{SaYo:pp} is written in terms of the Brinkmann
coordinates. Therefore we
need to take account of the coordinate transformation between two
coordinate systems. As we will see in Appendix, the coordinate
transformation does not affect our result in this paper.  

On the other hand, 
when the $0$-direction ($10$-direction) in the coordinate system
of the AdS$_4\times S^7$ (AdS$_7\times S^4$) is the
Dirichlet direction, then $\theta_+=0$ is obtained by the Penrose limit.
This means $\Gamma^{-+}\theta=\theta$, or equivalently $\Gamma^{12\cdots
9}\theta=\theta$.  Because $p$ is restricted to be 1, 5 and 9, the
matrix $M^{10-p}$ cannot be $\Gamma^{-+}$.  Thus we find ${\rm
M}^9=\Gamma^{12\cdots 9}$, which is the $(+-)$-brane boundary condition.
These two types of the resulting branes are exactly identical with those
found in \cite{SaYo:pp}.

\subsubsection*{Penrose Limit of M5-branes}

In the same manner used in the
study of 9-branes, the boundary conditions for the M5-branes are
modified in the Penrose limit according to the boundary conditions for the 
$0$ and $10$-directions.  

We can consider the (3,3)- and (1,5)-type M5-branes at the origin. 
First, the boundary condition for
$(3,3)$-brane, $\Gamma^{ a a_1'\cdots a_4'}\theta=\theta$, leads to
\begin{eqnarray}
\Gamma^{II_1'\cdots I_4'}\theta=\theta&~~~~~&\mbox{when $0,10$ $\in$ Neumann}
\,, \nn \\
\Gamma^{+-I_1'\cdots I_3'}\theta=\theta&~~~~~&\mbox{when $0,10$ $\in$ Dirichlet}\,,
\nn \\
\Gamma^{1\cdots 9}\theta=\theta&~~~~~&\mbox{otherwise}.
\end{eqnarray}
The last condition follows because in this case the boundary condition
becomes $\theta_+=0$ by taking the Penrose limit.
These are the boundary conditions for $(+,-;2,2)$-, $(3,3)$- and $(+,-)$-branes
respectively. Secondly, the boundary condition for $(1,5)$-brane,
$s\Gamma^{ a_1\cdots  a_3 a'b'}\theta=\theta$,
becomes
\begin{eqnarray}
\Gamma^{I_1\cdots I_3I'J'}\theta=\theta&~~~~~&\mbox{when $0, 10 \in$ Neumann}
\,, \nn \\
i\Gamma^{+-IJI'}\theta=\theta&~~~~~&\mbox{when $0, 10 \in$ Dirichlet}\,,\nn \\
\Gamma^{1\cdots 9}\theta=\theta&~~~~~&\mbox{otherwise}.
\end{eqnarray}
These are $(+,-;0,4)$-, $(1,5)$- and $(+,-)$-branes, respectively.
All of branes obtained in the Penrose limit are contained in the list
of the classification in the case of pp-wave \cite{SaYo:pp}.

\subsubsection*{Penrose Limit of 1-branes}

In this case, the boundary condition,
$s\Gamma^{a_1\cdots a_3a_1'\cdots a_6'}\theta=\theta$,
reduces to
\begin{eqnarray}
i\Gamma^{+-IJI'_1\cdots I_5'}\theta=\theta&~~~~~&
\mbox{when $0,10$ $\in$ Dirichlet}\,, \nn \\
\Gamma^{1\cdots 9}\theta=\theta&~~~~~& \mbox{otherwise},
\end{eqnarray}
which correspond to $(1,1)$- and $(+,-)$-branes, respectively.
Again, these are contained in the classification in \cite{SaYo:pp}.

\vspace*{0.5cm}

\begin{figure}
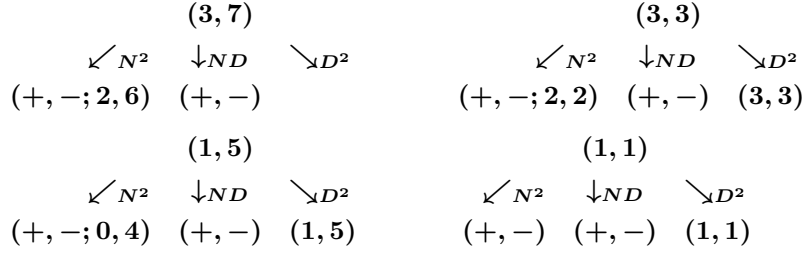

{\boldmath\footnotesize
\begin{eqnarray}
&&
  \begin{array}{rcl}
       & (3,7)   &    \\
       \swarrow_{N^2} &\downarrow_{ND} & \searrow_{D^2}    \\
     (+,-;2,6)  & (+,-)   & ~~~~\\
  \end{array}
~~~~~~~~~
  \begin{array}{rcl}
       & (3,3)   &    \\
       \swarrow_{N^2} &\downarrow_{ND} & \searrow_{D^2}    \\
     (+,-;2,2)  & (+,-)   & (3,3)\\
  \end{array}
\nonumber\\&&
  \begin{array}{rcl}
       & (1,5)   &    \\
       \swarrow_{N^2} &\downarrow_{ND} & \searrow_{D^2}    \\
     (+,-;0,4)  & (+,-)   & (1,5)\\
  \end{array}
~~~~~~~~~
  \begin{array}{rcl}
       & (1,1)   &    \\
       \swarrow_{N^2} &\downarrow_{ND} & \searrow_{D^2}    \\
     (+,-)  & (+,-)   & (1,1)\\
  \end{array}
\nonumber
\end{eqnarray}
}
\caption{Relation of Dirichlet branes in the AdS$_{4/7} \times S^{7/4}$
 and pp-wave background: 
\footnotesize Dirichlet branes in the AdS$_{4/7} \times S^{7/4}$ 
are mapped to those on the pp-wave background as shown above. Notably,
some 9-branes and M5-branes are mapped to the 1-branes expanding in 
 $(+,-)$-directions on the pp-wave background, when we choose a Neumann
 coordinate and a Dirichlet one in order to make the light-cone
 directions.}
\label{Penrose:fig}
\end{figure}

As a summary, we depict in Fig.\,\ref{Penrose:fig} the relation 
of branes in the AdS background to those in the pp-wave.
We should remark that all of the possible D-brane configurations 
in the pp-wave case have been recovered as the Penrose limit of 
those in the AdS case. Moreover, our result is consistent to 
the Penrose limit of the embedded 
AdS brane in eleven dimensions \cite{Kim-Yee}. 

Finally, we would like to comment on the Penrose limit of the 
Dirichlet branes outside the origin. 
As a matter of course, we can find the 1/4 BPS
Dirichlet branes even outside the origin by imposing a further boundary 
condition on the spinor $\th$, and some of such
configurations would be mapped to the 1/4 BPS or 1/2 BPS 1-branes 
outside the origin on the pp-wave background by taking the Penrose limit.

\section{Conclusion and Discussion}

In this paper we have discussed the boundary surfaces (Dirichlet branes)
of an open supermembrane in the AdS$_{4/7}\times S^{7/4}$ backgrounds.
By analyzing the covariant Wess-Zumino term, we have classified the
allowed configurations of Dirichlet branes at the origin.  All of 1/2
BPS configurations are mapped to those preserving a half of
supersymmetries on the pp-wave background obtained in \cite{SaYo:pp} by
taking the Penrose limit. In particular, when we choose a Neumann
direction and a Dirichlet one to form the light-cone directions, the
9-branes and M5-branes are changed into the 1-branes spanning in the
($+,-$)-directions. 

In addition, we have discussed the boundary surfaces
outside the origin, and it was found that 
the 1/2 BPS configurations are not allowed. As a
matter of course, it would be possible to find less supersymmetric
Dirichlet brane configurations, such as 1/4 BPS objects. This is an
interesting future problem.
It is also important to clarify the preserved
supersymmetries in the Penrose limit for these Dirichlet brane
configurations in the AdS$_{4/7}\times S^{7/4}$ background.

It is one of the most interesting problems to consider the application
of our results for the study of the AdS$_{4/7}\times S^{7/4}$
correspondence, and to clarify the field contents in
the dual gauge theory.
  Similar analysis with this paper is applicable for the
AdS string case, and we can classify the possible configurations of
D-branes of the AdS string,
and discuss their Penrose limits
using \cite{HKS:10}.  If such a study could succeed, then we
possibly might study the AdS/CFT correspondence at the string
theoretical level without using the Penrose limit as far as concerns
D-branes of the AdS string.
We will study in this direction as a future work \cite{SaYo3}.

\vspace*{1cm}

\noindent 
{\bf\large Acknowledgments}
\vspace*{0.5cm}

We would like to thank Machiko Hatsuda and Katsuyuki Sugiyama
for useful discussion.

\newpage 

\appendix 

\noindent 
{\large\bf  Appendix}
\section*{Notation and Convention}
In this place we will summarize miscellaneous notation and 
convention used in this paper. 

\subsection*{Notation of Coordinates}

For the supermembrane in the eleven-dimensional curved space-time, 
we use the following notation of supercoordinates for its superspace:
\begin{eqnarray}
(X^M,\th^{\al})\,, \qquad M = (\mu, \mu')\,,
~~~~~~~~~~
\mu \in AdS_4 (S^4)\,,~~~~
\mu' \in S^7 (AdS_7)\,,~~
\nn
\end{eqnarray}
and the background metric is expressed by $G_{MN}$. 
The coordinates in the Lorentz frame is denoted by 
\begin{eqnarray}
(X^A,\th^{\bar{\al}})\,, \qquad A=
(a,a')\,,
~~
a=\left\{
  \begin{array}{l}
  0,1,2,3     \\
  10,1,2,3    \\
  \end{array}
\right.\,,~~
a'=\left\{
  \begin{array}{ll}
  4,...,9,10     &~~\mbox{for}~~ AdS_4\times S^7    \\
  0,4,...,9   &~~\mbox{for}~~ AdS_7\times S^4    \\
  \end{array}
\right.\,,
\nn
\end{eqnarray}
and its metric is described by $\eta_{AB} =
\mbox{diag}(-1,+1,\ldots,+1)$\,
with $\eta_{00}=-1$.
The light-cone coordinates are defined by 
$
 X^{\pm} \equiv \frac{1}{\sqrt{2}}(X^0 \pm X^{10}) 
$\,.

The membrane world-volume is three-dimensional and its coordinates
are parameterized by $(\tau, \sigma^1, \sigma^2)$\,. 
The induced metric on the world-volume is represented by $g_{ij}$\,. 

\subsection*{$SO(10,1)$ Clifford Algebra}

We denote a 32-component Majorana spinor as $\th$, and 
the $SO(10,1)$ gamma matrices $\Gamma^A$'s satisfy the 
$SO(10,1)$ Clifford algebra 
\begin{eqnarray}
 & & \{\Gamma^A,\,\Gamma^B\} = 2\eta^{AB}\,, 
\quad  \{\Gamma^M,\,\Gamma^N \} = 2G^{MN}
\,, \quad \Gamma^A \equiv e^A_M\Gamma^M\,, 
\quad \Gamma^M \equiv  e^M_A\Gamma^A\,,
\nn 
\end{eqnarray}
where the light-cone components of the $SO(10,1)$ gamma matrices are 
\begin{eqnarray}
&&\Gamma^{\pm} \equiv \frac{1}{\sqrt{2}}\left(\Gamma^0 \pm \Gamma^{10}\right),
 \quad \{ \Gamma^+,\, \Gamma^- \} = -2 \mathbb I_{32}\,.
\nn 
\end{eqnarray}
We shall choose $\Gamma^A$'s such that $\Gamma^0$ 
is anti-hermite matrix and others 
are hermite matrices.  In this choice the relation 
$(\Gamma^A)^{\dagger} = \Gamma_0\Gamma^A\Gamma_0$ 
is satisfied. 
The charge conjugation of $\th$ 
is defined by 
\begin{equation}
 \th^{\s C} \equiv \mathcal{C}\bar{\th}^{\s T}\,, \nn 
\end{equation}
where $\bar{\th}$ is the Dirac conjugation of $\th$ and is defined by 
$\bar{\th} \equiv \th^{\dagger}\Gamma_0$. The charge conjugation matrix 
$\mathcal{C}$ satisfies the following relation:
\begin{eqnarray}
(\Gamma^A)^{\s T} = - \mathcal{C}^{-1}\Gamma^A \mathcal{C}\,, 
\quad \mathcal{C}^{\s T} = - \mathcal{C}\,. \nn
\end{eqnarray}
For an arbitrary Majorana spinor $\th$ satisfying the 
Majorana condition $\th^{\s C}=\th$, we can easily show the formula
\begin{equation}
\bar{\th} = - \th^{\s T}\mathcal{C}^{-1}\,. \nn 
\end{equation}
That is, the charge conjugation matrix $\mathcal{C}$ is defined by
$\mathcal{C}=\Gamma_0$ in this representation. The $\Gamma^A$'s are real
matrices (i.e., ($\Gamma^A)^{\ast} = \Gamma^A$). We also see that
$\Gamma^r$~($r=1,2,...,9$) and $\Gamma^{10}$ are symmetric and $\Gamma^{0}$ is
skewsymmetric.

\subsection*{Relation to the KG solution}

The Kowalski-Glikman solution, which is the maximally supersymmetric 
pp-wave background in eleven dimensions, can be obtained 
from AdS$_{4/7}\times S^{7/4}$ by taking the Penrose limit.
Noting that the super-isometry algebra of 
the KG solution is an In\"on\"u-Wigner contraction
of that of AdS$_{4/7}\times S^{7/4}$ \cite{HKS:11},
the group manifold for the pp-wave algebra
is parameterized by $g={\rm e}^{X^+P_++X^-P_-+X^mP_m}$
because we used the parameterization $g={\rm e}^{X^aP_a+X^{a'}P_{a'}}$
in (2.5).
The metric in this parametrization
is written as
\begin{eqnarray*}
ds^2&=&-2e^+e^-+(e^I)^2+(e^{I'})\,,
\end{eqnarray*}
where the vielbeins are given by
\begin{eqnarray}
e^+&=&dX^+\,, \quad z=\frac{\mu}{3}{X}^+\,, \quad
 z'=\frac{\mu}{6}{X}^+\,, \nn \\
e^-&=&dX^-
-\left(1-\frac{\sin z}{z}\right)
\frac{X^I}{(X^+)^2}
(X^+dX^I-X^IdX^+) \nn \\
&& \quad -\left(1-\frac{\sin z'}{z'}\right)
\frac{X^{I'}}{(X^+)^2}
(X^+dX^{I'}
-X^{I'}dX^+)\,, \nn \\
e^I &=& 
\frac{\sin z}{z}dX^I+\left(1-\frac{\sin z}{z}\right)\frac{X^I}{X^+}dX^+\,,
\quad 
e^{I'}=
\frac{\sin z'}{z'}dX^{I'}+\left(1-\frac{\sin z'}{z'}\right)
\frac{X^{I'}}{X^+}dX^+\,. \nn 
\end{eqnarray}
In order to make the metric to be of the ordinary form
in Brinkmann coordinate
we change the coordinate $(X^\pm,X^m)$ as
\begin{eqnarray}
 \hat{X}^+ &\equiv& X^+\, , \quad \hat X^{I} \;\equiv\; 
\frac{\sin z}{z}X^{I}~, \quad 
\hat X^{I'} \;\equiv\; \frac{\sin z'}{z'}X^{I'}\,,  \nn  \\
\hat X^- &\equiv& X^- 
-\frac{\mu}{6}\frac{z-\sin z\cos z}
 {\sin^2z}(\hat X^I)^2
-\frac{\mu}{12}\frac{z'-\sin z'\cos z'}
 {\sin^2 z'}(\hat X^{I'})^2\,. \nn 
\end{eqnarray}
Under
this transformation, the metric and the flux become
\begin{eqnarray}
ds^2 &=& 
-2d\hat{X}^+d\hat X^-
-\Bigl[
\left(\frac{\mu}{3}\right)^2(\hat X^i)^2 + 
\left(\frac{\mu}{6}\right)^2(\hat X^{i'})^2 
\Bigr](d\hat{X}^+)^2
+(d\hat X^m)^2\,, \nn 
\end{eqnarray}
and the constant flux $F_{+123} = \mu$, respectively. 
This reveals the fact that
$X^i=0$ and $X^{i'}=0$ for AdS$_{4/7}\times S^{7/4}$ correspond to $\hat
X^i=0$ and $\hat X^{i'}=0$ for the KG solution in Brinkmann
coordinates. We can easily see that the condition $X^+=X^- = 0$ 
corresponds to $\hat{X}^+ = \hat{X}^- =0$ by the use of the formulae of
trigonometrical functions.

\end{document}